\documentclass[shortnote]{jpsj2} %% for short notes

\usepackage{graphics}

\title{Generalized Cubic Model for BaTiO$_3$-like Ferroelectric Substance}

\author{Norikazu \textsc{TODOROKI}$^{1}$\thanks{E-mail
address: todoroki@kanagawa-u.ac.jp} }

\inst{$^{1}$
Institute of Physics, Kanagawa University, 3-27-1 Rokkaku-bashi, 
Kanagawa-ku, Yokohama 221-8686
}

\kword{%
BaTiO$_3$, ferroelectric substance, order-disorder model, structural phase transition, incompletely ordered phase
}

\begin{document}
\sloppy
\maketitle
Barium titanate BaTiO$_3$ (BTO) exhibits a large electric polarization at room temperature and has a high piezoelectric coefficient. 
Therefore, it is an important compound used in various applications among substances with a perovskite structure. 
BTO undergoes three successive phase transitions at $T\simeq 403$~K, $278$~K and $183$~K from the cubic paraelectric phase 
at high temperature to three ferroelectric phases at low temperature; 
the three phases exhibit tetragonal, orthorhombic, and rhombohedral symmetries.
Many researchers have studied the fundamental properties and applications of BTO since its discovery. \cite{shirane}
The complicated order process attracts interest of many theoretical physicists.
However, the property of the ordered phases and the process of ordering of BTO are not sufficiently comprehended. 
To comprehend the ordering process, several studies were carried out to determine whether this substance belongs to 
the order-disorder or the displacive type. 
In a recent numerical study, it was suggested that BTO was not described as a simple displacive model and 
had order-disorder character. \cite{Zhong}

If this substance is of the order-disorder type, we expect that the ordering process mechanisms describe an order-disorder model as a spin model.
The order-disorder model of this substance was proposed  for the first time by Mason and Matthias. \cite{Mason}
Their model was very simple and was incapable of explaining the phase transitions and ordered phases of BTO.
Subsequently, many models were proposed, but were incapable of understanding the mechanisms of the ordering process of BTO.
In particular, Inoue's explanation of his experimental results using the Yamada model was of interest. 
The Yamada model quantitatively was in good quantitative agreement with the experimental results.
 \cite{Inoue1}

However, the existence of some randomness in intermediate phase is reported by an X-ray analysis. \cite{Comes} 
In addition, an experimental result of hyper-Raman scattering suggests 
the existence of displaced clusters at a higher temperature than the cubic-tetragonal phase transition temperature 
in the [111] directions. \cite{Inoue2}
The previous models that include the Yamada model cannot describe the nature of the ordered phases. 
On the other hand, most of the previous models are used to describe only cubic-tetragonal transition.
It is expected that a model describing BTO over the entire temperature region will be developed.

Therefore, the aims of the present study are to understand the phase transitions and ordered phases by using the order-disorder type microscopic model. 
We qualitatively describe all the ordered phases and phase transitions by our model very well. 

We consider the following Hamiltonian:
\begin{eqnarray}
{\cal H}=\sum_{\langle i, j\rangle}\epsilon_{n_i,n_j}.
\end{eqnarray}
Here, the summation $\sum_{\langle i, j\rangle}$ runs over the nearest-neighbor pairs of a simple cubic lattice, 
$n_i$ is the spin variable that can assume eight states as shown in Fig.1 and the energy parameter $\epsilon_{n,m}$ is 
\begin{eqnarray}
[\epsilon_{n,m}]=\left [ 
\begin{array}{cccccccc}
0 & \alpha+\delta_y & \beta+\delta_{xy} & \alpha+\delta_x & \alpha+\delta_z & \beta+\delta_{yz} & 1 & \beta+\delta_{zx} \\
\alpha+\delta_y & 0 & \alpha+\delta_x & \beta+\delta_{xy} & \beta+\delta_{yz} & \alpha+\delta_z & \beta+\delta_{zx} & 1 \\
\beta+\delta_{xy} & \alpha+\delta_x & 0 & \alpha+\delta_y & 1 & \beta+\delta_{zx} & \alpha+\delta_z & \beta+\delta_{yz} \\
\alpha+\delta_x & \beta+\delta_{xy} & \alpha+\delta_y & 0 & \beta+\delta_{zx} & 1 & \beta+\delta_{yz} & \alpha+\delta_z \\
\alpha+\delta_z & \beta+\delta_{yz} & 1 & \beta+\delta_{zx} & 0 & \alpha+\delta_y & \beta+\delta_{xy} & \alpha+\delta_x \\
\beta+\delta_{yz} & \alpha+\delta_z & \beta+\delta_{zx} & 1 & \alpha+\delta_y & 0 & \alpha+\delta_x & \beta+\delta_{xy} \\
1 & \beta+\delta_{zx} & \alpha+\delta_z & \beta+\delta_{yz} & \beta+\delta_{xy} & \alpha+\delta_x & 0 & \alpha+\delta_y \\
\beta+\delta_{zx} & 1 & \beta+\delta_{yz} & \alpha+\delta_z & \alpha+\delta_x & \beta+\delta_{xy} & \alpha+\delta_y & 0
\end{array}
\right ],
\end{eqnarray}
with
\begin{eqnarray}
\delta_x&=&\delta (P_x^2-P_y^2-P_z^2),\\
\delta_y&=&\delta (P_y^2-P_z^2-P_x^2),\\
\delta_z&=&\delta (P_z^2-P_x^2-P_y^2),\\
\delta_{xy}&=&\delta (P_x^2+P_y^2-P_z^2),\\
\delta_{yz}&=&\delta (P_y^2+P_z^2-P_x^2),\\
\delta_{zx}&=&\delta (P_z^2+P_x^2-P_y^2),
\end{eqnarray}
where $P_i$ is defined by the density of the $i$-th state $p(i)$ as follows
\begin{eqnarray}
P_x&=&(p(1)+p(2)+p(5)+p(6))-(p(3)+p(4)+p(7)+p(8)), \\
P_y&=&(p(2)+p(3)+p(6)+p(7))-(p(1)+p(4)+p(5)+p(8)), \\
P_z&=&(p(5)+p(6)+p(7)+p(8))-(p(1)+p(2)+p(3)+p(4)).
\end{eqnarray}
$P_i$ does not represent polarization
because our model is simplified, but this corresponds to the polarization qualitatively.
In Fig.\ref{fig2}, we show the energy level of a neighboring spin pair with $\delta=0$.
$\Delta \theta$ denotes a relative angle between the neighboring spins.
The energy of the neighboring spin pair is $\alpha$ with $\Delta \theta=2\tan^{-1}(1/\sqrt{2})$
and $\beta$ with $\Delta \theta=2\tan^{-1}(\sqrt{2})$, namely, 
 the energy is determined by the relative angle such as in the generalized clock model. \cite{Ueno,Todoroki1}
The parameter $\delta$ denotes an effect of the relative angles change by a lattice distortion. 
The energy levels modulate by $\delta$ on the polarized phase.
This model exhibits cubic symmetry as indicated by $\delta\neq 0$ as well as $\delta= 0$.

We treat this model by a pair approximation 
%%The self consistent equations are
%%\begin{eqnarray}
%%x_{m,n}=\exp\left ( 2\beta\lambda/z-\beta\epsilon_{m,n}\right ) (p(m)p(n))^{1-1/z},
%%\end{eqnarray}
%%\begin{eqnarray}
%%\exp \left ( -2\beta\lambda/z \right )=\sum_m\sum_n \exp(-\beta\epsilon_{m,n})(p(m)p(n))^{1-1/z},
%%\end{eqnarray}
%%where $x_{m,n}$ is the probability that 
%%a pair of neigboring spins are in the $m$-th and $n$-th states.
%%$x_{m,n}$ satisfy following relations.
%%\begin{eqnarray}
%%\sum_m\sum_n x_{m,n}=1,\hspace{1cm} p_m=\sum_nx_{m,n}, \hspace{1cm} x_{m,n}=x_{n,m}.
%%\end{eqnarray}
and solve these self-consistent equations numerically. \cite{Kikuchi,Ueno}
In Fig.3, we show the density of the each state at $\alpha=0.2$, $\beta=0.4$, and $\delta=0.04$.
These values of parameters are chosen so that the ratios of the transition temperatures accorded with an experiment results approximately. 
However, it must be noted that those values do not have a meaning quantitatively because our model is simplified.
We observe three first-order phase transitions and four phases. 
We denote these phases as phase I, phase II, phase III, and phase IV from the high temperature side.
Phase I is a paraelectric phase with cubic symmetry. 
Phase II, with tetragonal symmetry, has four states sharing one face among eight states indicating mixing. 
Phase III, with orthorhombic symmetry, has two states sharing one edge among eight states indicating mixing. 
In phase IV, with trigonal symmetry, 
the system completely orders one state among eight states.
In Fig.\ref{fig4}, we show the relationship between the mixing states and polarization direction on each phase.
These results are in good agreement with an experimental result.
The intermediate phases correspond to the incompletely ordered phase (IOP) in the generalized clock model \cite{Todoroki1} 
or partially disordered phase in a stacked triangular Ising model. \cite{Mekata,Todoroki2} 
When $\alpha$, which is the energy between the neighboring states, is less than the temperature, 
it is possible that these intermediate phases are realized for entropy gain by the mixing of the neighboring states. 

When the temperature is higher than the cubic-tetragonal transition temperature, the eight states are mixed and the domains of each state grow up to a specific size 
for the correlation with temperature nearing the transition point.
Therefore, it is likely that the displaced clusters are detected by hyper-Raman scattering.
On intermediate phases II and III, each of the four and two states constitute clusters of limited size; further, they are mixed.
It is possible that the result of X-ray analysis is described by the above mentioned cluster.

In summary, our model has three phase transitions and four phases.
The symmetry and directions of the polarizations 
of the ordered phases agree with the experimental results of BTO.
The unexplained experimental results of the intermediate phases by X-ray analysis and hyper-Raman scattering may be explained by our model.
The intermediate phases (phases II and III) are considered as a type of IOP.
It may be noted that the process of ordering the same as that was suggested by Takahashi. \cite{Takahashi}
Although our model quantitatively describes the phase transitions and ordered phases, a quantitative argument is necessary for the complete understanding of BTO.
Therefore, it may be necessary to consider the spin-phonon system such as in the Yamada model. 
Hence, there is scope for further investigation. 

%%\section*{Acknowledgement}
I would like to thank S. Tanaka and Prof. I. Oonari and Dr. H. Watanabe for 
their helpful discussions.
I also thank J. Harada for her tremendous support.

\newpage

\begin{figure}
\includegraphics[scale=0.5]{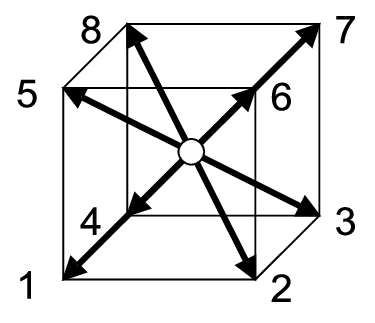}
\caption[]{States of the spin variables with a cubic symmetry. 
Each spin assumes eight states corresponding to the vertex of a cube.\label{fig1}}
\end{figure}

\begin{figure}
\includegraphics[scale=0.5]{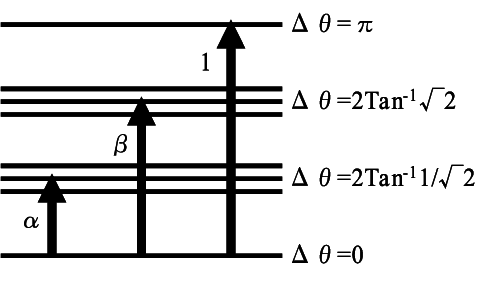}
\caption[]{Energy level of a neighboring spin pair with $\delta=0$.
The first and second excitation states have a three-fold degeneration for symmetry.
\label{fig2}}
\end{figure}

\begin{figure}
\includegraphics[scale=0.5]{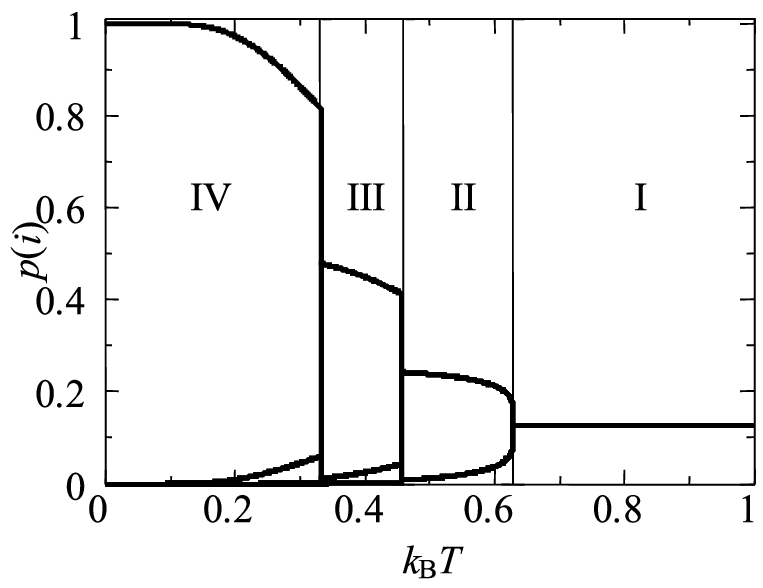}
\caption[]{Temperature dependence density of the each state at $\alpha=0.2$, $\beta=0.4$, and $\delta=0.04$.
There are two intermediate phases except the paraelectric phase (phase I) and completely ordered phase (phase IV). 
On the high temperature side intermediate phase (phase II), the four states that share one face are dominantly well mixed. 
On the low temperature side intermediate phase (phase III), the two states that share one edge are dominantly well mixed. 
\label{fig3}}
\end{figure}

\begin{figure}
\includegraphics[scale=0.5]{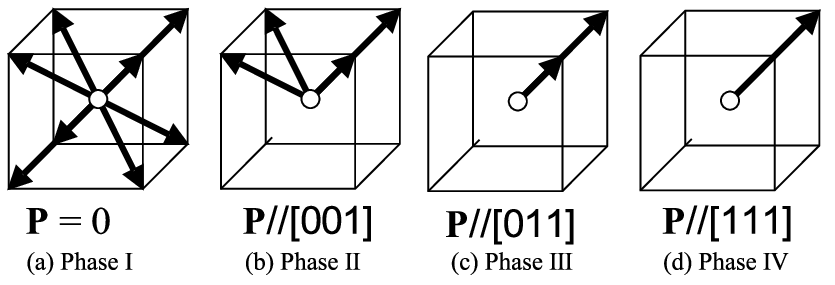}
\caption[]{The schematic figure 
of the relationship between the mixing states and polarization direction of each phase. 
\label{fig4}}
\end{figure}
\end{document}